# Leveraging Microgrids for Capturing Uncertain Distribution Network Net Load Ramping

Alireza Majzoobi, Amin Khodaei
Dept. of Electrical and Computer Engineering
University of Denver
Denver, CO, USA
Alireza.Majzoobi@du.edu, Amin.Khodaei@du.edu

*Abstract—* **In this paper, a flexibility-oriented microgrid optimal scheduling model is proposed to mitigate distribution network net load variability caused by large penetration distributed solar generation. The distributed solar generation variability, which is caused by increasing adoption of this technology by end-use consumers, is mainly addressed by electric utilities using grid reinforcement. Microgrids, however, provide viable and local solutions to this pressing challenge. The proposed model, which is developed using mixed-integer programming and employs robust optimization, not only can efficiently capture distribution network net load variations, mainly in terms of ramping, but also accounts for possible uncertainties in forecasting. Numerical simulations on a test distribution feeder with one microgrid and several consumers/prosumers indicate the effectiveness of the proposed model.**

*Keywords—Microgrid, optimal scheduling, ramping, renewable energy, uncertainty, utility grid support*

## Nomenclature

*Indices:*

| | |
|---|---|
| $c$ | Superscript for distribution network consumers and prosumers. |
| ch/dch | Superscript for energy storage charging/discharging mode. |
| $d$ | Index for loads. |
| $i$ | Index for DERs. |
| $j$ | Index for consumers/prosumers at the distribution network. |
| $t$ | Index for time periods (hour). |
| $u$ | Superscript for the utility grid. |

*Sets:*

| | |
|---|---|
| D | Set of adjustable loads. |
| G | Set of dispatchable units. |
| N | Set of consumers/prosumers. |
| P | Set of primal variables. |
| S | Set of energy storage systems. |
| U | Set of uncertain parameters. |

*Parameters:*

| | |
|---|---|
| DR/UR | Ramp down/up rate. |
| DT/UT | Minimum down/up time. |
| E | Load total required energy. |
| F(.) | Generation cost. |
| MC/MD | Minimum charging/discharging time. |
| MU | Minimum operating time. |
| $\alpha, \beta$ | Specified start and end times of adjustable loads. |
| $\rho^M$ | Market price. |
| $\eta$ | Energy storage efficiency. |
| $\tau$ | Time period. |

*Variables:*

| | |
|---|---|
| C | Energy storage available (stored) energy. |
| D | Load demand. |
| I | Commitment state of dispatchable units. |
| P | DER output power. |
| $P^M$ | Utility grid power exchange with the microgrid. |
| SD/SU | Shut down/startup cost. |
| $T^{ch}/T^{dch}$ | Number of successive charging/discharging hours. |
| $T^{on}/T^{off}$ | Number of successive ON/OFF hours. |
| $u$ | Energy storage discharging state (1 when discharging, 0 otherwise). |
| $v$ | Energy storage charging state (1 when charging, 0 otherwise). |
| $z$ | Adjustable load state (1 when operating, 0 otherwise). |

## I. Introduction

THE evolution of renewable energy over the past few decades has surpassed all expectations and the renewable energy technologies are rapidly growing throughout the world. Environmental concerns about climate change, especially in recent years, have been one of the key factors for the significant increase in renewable energy investment. The global investment in renewable energy in 2015 reached $285.9 billion, surpassing its 2011 record of $278.5 billion. In 2015, 53.6% of all installed and commissioned generation capacity was renewable energy (about 134 GW) [1]. In addition to air pollution reduction, reduced operation cost has been another reason for growth of renewable energy in recent years. In the last five years, the cost of energy from renewable sources such as solar and wind has dramatically dropped by 78% and 58%, respectively [2]. Today, even without subsidies and supports of governments, these sources of energy are cost competitive with conventional generation resources in many parts of the world [2].

Despite all advantages of renewable energy resources, there are some drawbacks which should be carefully taken into account. For instance, growing solar energy deployment as a favorable distributed energy resource and participation of consumers in decarbonized clean energy production, has changed the typical daily demand curves. Considering that the solar generation is highest around the noontime, the daily net load curve, i.e. the difference of load and solar generation, drops significantly at noon and increases in early evening hours due to sunset and residential load increase. Therefore, the utility companies encounter a sharp ramping in daily demand curves. The California Independent System Operator's report on this changing demand profile, published in 2013, predicted as high as 4.3 GW/h required ramping in daily demand curve by 2020 [3]. Furthermore, unlike the conventional energy resources, renewable generation is inherently variable (generation constantly varies) and uncertain (generation cannot be forecasted with perfect accuracy) [4]. The studies by the North American Electric Reliability Corporation (NERC) show that the power generation of solar panels can change by ±70%, in a timeframe of 2–10 minutes, several times per day [4], in addition to the typical ±1% to ±7% deviation between predicted demand and actual demand in the system [5].

The high penetration of renewable generation has significantly increased the system uncertainty which further challenges traditional methods in cost-effective and reliable control, operation, and planning of power systems [6], [7]. Several methods are proposed to address this challenge [8]-[10]. The traditional solution of utility companies for maintaining system supply-demand balance is to utilize fast ramping units that can be quickly dispatched and ramped up/down. With high penetration of renewable energy resources, the spinning reserve of these units should be increased which imposes higher operation cost to the utility companies and decreases the efficiency of the system. Energy storage can also be used to capture the renewable generation uncertainty [11]-[13]. Energy storage, however, is still an expensive technology and its large-scale deployments are limited. Demand response, as a way to modify electricity consumption to increase power system efficiency and reliability, is another proposed solution. The successful implementation of the demand response, however, needs extensive infrastructures as well as considerable participation of consumers [14]-[17]. Sun tracking and rotating solar panels can be considered as other existing methods which are usually more applicable in solar farms than distributed applications [18].

Leveraging potential flexibility of existing microgrids in distribution networks as a local, novel, and viable method has been proposed in [19] and extended in this paper to address aforementioned challenges and to alleviate the negative impacts of increasing renewable penetration. Microgrids, as small-scale power system with the ability of operating in both grid-connected and islanding modes [20], have attracted considerable attention in recent years, primarily due their promising features in enhancing reliability, resilience, and power quality, reducing environmental impact, relieving network congestion, and improving energy efficiency [21]-[29]. Between 2011 and 2014, more than $213 million has been invested on microgrid projects in the United States [30]. Microgrids' deployment of dispatchable generation units, energy storage, and adjustable loads, provides significant controllable fast-response generation that can be used for flexibility and ramping purposes, as will be discussed and modeled in this paper. In addition, the existing uncertainty in distributed load, distributed solar generation, and price will be captured by the microgrid.

The remaining parts of the paper are organized as follows. The outline of the proposed model is explained in Section II. Section II further presents model formulations including max-min objective function and microgrid operation and flexibility constraints. The effectiveness of the proposed model is proved in Section III, via numerical simulations. Discussion on the results and features of the proposed model are also presented in this section. The conclusions are provided in Section IV.

## II. PROBLEM MODELING AND FORMULATION

### A. Problem Statement

The power ($P^u$) that the electric utility should supply to a certain distribution feeder is equal to the microgrid net load ($P^M$) plus the aggregated net load of other customers, including consumers and prosumers, in this feeder ($P^c$) as presented in (1).

$$P_t^u = P_t^M + \sum_{j \in N} P_{jt}^c \qquad \forall t. \qquad (1)$$

The net load of consumers and prosumers is highly variable and uncertain, primarily due to the deployment of distributed renewable energy resources, and is further uncontrollable from the utility side. The net load of the microgrid, moreover, is controlled by the microgrid controller based on economy and reliability considerations. The summation of these two uncontrollable net loads with considerable levels of renewable generation causes variability (mainly in terms of large ramps) and uncertainty for the power that the electric utility needs to provide. A viable solution, however, is to incentivize the microgrid to locally capture the ramping, i.e., not only the microgrid retracts its variability, but also helps the electric utility in capturing the variability of other customers connected to the same distribution feeder. To model this, the utility grid ramping limit (2) should be translated into proper limitations on the microgrid net load as discussed in the next subsection.

$$\left| P_t^u - P_{(t-1)}^u \right| \leq \Delta \qquad \forall t. \qquad (2)$$

### B. Problem Formulation

The flexibility-oriented microgrid optimal scheduling model under uncertainty is proposed as in (3)-(21).

$$\max_U \min_P \sum_t [\sum_{i \in G} F(P_{it}) + \rho_t^M P_t^M] \qquad (3)$$

$$\sum_i P_{it} + P_t^M = \sum_d D_{dt} \qquad \forall t, \qquad (4)$$

$$-P^{M,\max} \le P_t^M \le P^{M,\max} \qquad \forall t, \qquad (5)$$

$$P_i^{\min} I_{it} \le P_{it} \le P_i^{\max} I_{it} \qquad \forall i \in G, \forall t, \qquad (6)$$

$$P_{it} - P_{i(t-1)} \le UR_i \qquad \forall i \in G, \forall t, \qquad (7)$$

$$P_{i(t-1)} - P_{it} \le DR_i \qquad \forall i \in G, \forall t, \qquad (8)$$

$$T_{it}^{on} \ge UT_i(I_{it} - I_{i(t-1)}) \qquad \forall i \in G, \forall t, \qquad (9)$$

$$T_{it}^{off} \ge DT_i(I_{i(t-1)} - I_{it}) \qquad \forall i \in G, \forall t, \qquad (10)$$

$$P_{it} \le P_{it}^{dch,\max} u_{it} - P_{it}^{ch,\min} v_{it} \qquad \forall i \in S, \forall t, \qquad (11)$$

$$P_{it} \ge P_{it}^{dch,\min} u_{it} - P_{it}^{ch,\max} v_{it} \qquad \forall i \in S, \forall t, \qquad (12)$$

$$u_{it} + v_{it} \le 1 \qquad \forall i \in S, \forall t, \qquad (13)$$

$$C_{it} = C_{i(t-1)} - (P_{it} u_{it} \tau / \eta_i) - P_{it} v_{it} \tau \qquad \forall i \in S, \forall t, \qquad (14)$$

$$C_i^{\min} \le C_{it} \le C_i^{\max} \qquad \forall i \in S, \forall t, \qquad (15)$$

$$T_{it}^{ch} \ge MC_i(u_{it} - u_{i(t-1)}) \qquad \forall i \in S, \forall t, \qquad (16)$$

$$T_{it}^{dch} \ge MD_i(v_{it} - v_{i(t-1)}) \qquad \forall i \in S, \forall t, \qquad (17)$$

$$D_{dt}^{\min} z_{dt} \le D_{dt} \le D_{dt}^{\max} z_{dt} \qquad \forall d \in D, \forall t, \qquad (18)$$

$$T_d^{on} \ge MU_d(z_{dt} - z_{d(t-1)}) \qquad \forall d \in D, \forall t, \qquad (19)$$

$$\sum_{t \in [\alpha_d, \beta_d]} D_{dt} = E_d \qquad \forall d \in D, \qquad (20)$$

$$\Delta_t^{low} \le P_t^M - P_{(t-1)}^M \le \Delta_t^{up} \qquad \forall t. \qquad (21)$$

The objective of this problem is to minimize the microgrid operation cost over primary variables and to maximize over uncertain variables. The microgrid operation cost consists of two terms; local generation cost, i.e. the first term of (3), and the cost of power exchange with the utility grid, i.e. the second term of (3). It should be noted that the primary variables are local distributed energy resources (DERs), loads, and utility grid power exchange, while uncertain variables are the net load of aggregated customers and the electricity price. Thus, a robust solution (i.e., the worst-case) will be calculated which ensures that the microgrid can capture distribution network net load ramping even if load, generation, and price forecasts are uncertain.

This objective is subject to system constraints (4)-(5), component constraints (6)-(20), and the flexibility constraint (21). The load balance equation (4) ensures that adequate generation is available (locally and purchased from the utility grid) to supply local loads. The capacity of the line connecting the microgrid to the utility grid defines the restriction on the exchanged power (5). Dispatchable units are subject to minimum and maximum generation capacity limits (6), ramping limits (7)-(8), and minimum up/down time limits (9)-(10). Constraints (11)-(17) define the restrictions on energy storage. The maximum and minimum amounts of charging and discharging are define by (11) and (12). Constraint (13) checks the energy storage operation mode to ensure that it does not operate at both charging and discharging modes simultaneously. Available energy at each hour is calculated with (14), while its limitations are defined in (15). Constraints (16) and (17) specify the minimum charging and discharging time limits, respectively. Constraints (18)-(20) define the restrictions on adjustable loads, including rated power limitations (18), the minimum operating time (19), and the required energy to complete an operating cycle (20) [24]. Constraint (21) is the utility ramping limit which is translated into a constraint on the microgrid net load. This constraint is obtained by substituting the value of the utility power from (1) in (2) and rearranging the terms. The lower and upper limits, which now are functions of time, are calculated based on the net load of connected customers as in (22) and (23):

$$\Delta_t^{low} = -\Delta - (\sum_j P_{jt}^c - \sum_j P_{j(t-1)}^c) \qquad \forall t, \qquad (22)$$

$$\Delta_t^{up} = \Delta - (\sum_j P_{jt}^c - \sum_j P_{j(t-1)}^c) \qquad \forall t. \qquad (23)$$

C. Solution Approach

The proposed robust model is decomposed into a master problem and a subproblem using Benders decomposition as illustrated in Fig. 1. The master problem calculates the minimum operation cost considering only constraints that include binary variables. It can be represented as follows:

$$\min_P \sum_t \sum_{i \in G} F(P_{it}) \qquad (24)$$

subject to (6), (9)-(13), (16)-(19).

Once binary scheduling variables are determined, including the DERs and loads schedules, these variables are sent to the subproblem, defined as follows:

$$\max_U \min_P \sum_t \rho_t^M P_t^M \qquad (25)$$

subject to (4)-(8), (11)-(12), (14)-(15), (18), (20)-(21), and given binary variables from the master problem.

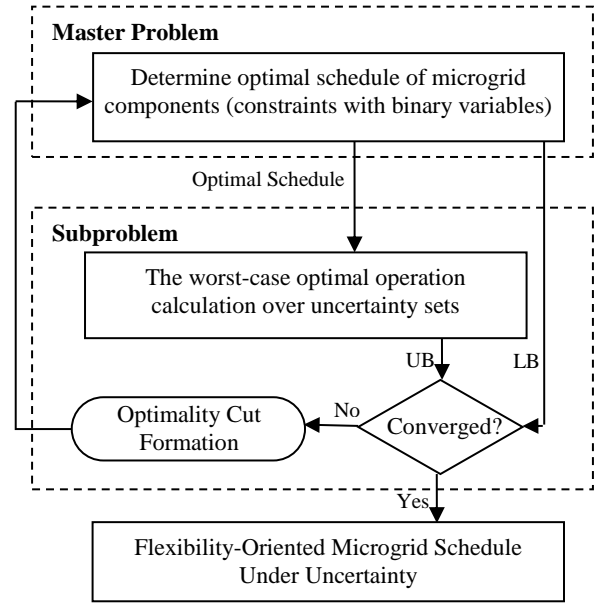

Fig. 1 Flowchart of the proposed flexibility-oriented microgrid optimal scheduling model.

The subproblem finds the microgrid's worst-case minimum operation cost over uncertainty sets based on the fixed schedules from the master problem. Since there is no binary variable in the subproblem, it is possible to convert the inner minimization problem into a maximization problem via duality theory and further combine the two maximization problems. Each uncertain parameter varies in an interval which is obtained from the forecasted value and expanded around the forecasted value based on the forecast error (i.e., a polyhedral uncertainty set). The robust optimization method finds the worst-case optimal operation solution while uncertain parameters vary within their associated uncertainty intervals. In order to control the robustness and restrict the solution conservatism, the budget of uncertainty is defined for confining the numbers of uncertain parameters which can take their worst-case values [21].

This robust optimization approach integrates uncertainties of distributed load, distributed solar generation, and market price forecasts. Once solved, the optimal DERs and loads schedules will be obtained which also ensure flexibility. Checking the lower and upper bound proximity of the problem is an approach for examining the solution convergence. As it is shown in the Fig. 1, the lower and the upper bounds of the problem are calculated in the master problem and the subproblem, respectively. The optimality cut will be formed in the subproblem and sent back to the master problem for updating the current schedule, if the solution is not converged. This iterative process continues until the convergence criterion is met and the solution is proven optimal [21].

### III. NUMERICAL EXAMPLES

The microgrid used for studying the performance of the proposed model in this paper consists of two nondispatchable units (solar and wind), four dispatchable units, one energy storage, and five adjustable loads. The characteristics of these energy resources, loads, as well as the hourly market price are available in [22]. The amount of aggregated load and solar generation in distribution feeder are tabulated in Tables I and II, respectively. A 10 MW capacity is assumed for the line between the microgrid and the utility grid. The developed mixed-integer programming problems are solved using CPLEX 12.6.

TABLE I
AGGREGATED DISTRIBUTED LOAD

| Time (h) | 1 | 2 | 3 | 4 | 5 | 6 |
|---|---|---|---|---|---|---|
| Load (MW) | 13.50 | 12.50 | 11.80 | 11.70 | 12.10 | 12.50 |
| Time (h) | 7 | 8 | 9 | 10 | 11 | 12 |
| Load (MW) | 12.80 | 14.00 | 14.60 | 15.20 | 16.00 | 17.00 |
| Time (h) | 13 | 14 | 15 | 16 | 17 | 18 |
| Load (MW) | 18.50 | 18.00 | 17.00 | 16.70 | 17.00 | 18.00 |
| Time (h) | 19 | 20 | 21 | 22 | 23 | 24 |
| Load (MW) | 20.25 | 20.65 | 19.00 | 17.00 | 14.50 | 13.80 |

TABLE II
AGGREGATED DISTRIBUTED SOLAR GENERATION

| Time (h) | 1 | 2 | 3 | 4 | 5 | 6 |
|---|---|---|---|---|---|---|
| Power (MW) | 0.00 | 0.00 | 0.00 | 0.00 | 0.00 | 0.00 |
| Time (h) | 7 | 8 | 9 | 10 | 11 | 12 |
| Power (MW) | 0.00 | 0.00 | 1.00 | 4.00 | 8.00 | 11.50 |
| Time (h) | 13 | 14 | 15 | 16 | 17 | 18 |
| Power (MW) | 14.00 | 14.20 | 14.00 | 12.40 | 11.00 | 6.00 |
| Time (h) | 19 | 20 | 21 | 22 | 23 | 24 |
| Power (MW) | 2.75 | 0.85 | 0.00 | 0.00 | 0.00 | 0.00 |

Three cases are studied to show the effectiveness of the proposed model for addressing distribution network flexibility concerns:
**Case 1:** Flexibility-oriented microgrid optimal scheduling ignoring uncertainty.
**Case 2:** Flexibility-oriented microgrid optimal scheduling considering uncertainty.
**Case 3:** Sensitivity analysis with regards to the budget of uncertainty.

**Case 1:** The flexibility-oriented microgrid optimal scheduling without consideration of any uncertainty is solved for a 24-hour horizon. The microgrid should capture the rampings above desired amount of the utility grid, which has been assumed as 2 MW/h in this case. When there is no contribution from the microgrid, the utility grid should capture the ramping of distribution feeder net load, for instance a maximum of 6 MW/h load change or an average of 4.6 MW/h in 3 hours. In this condition, unit 1 is ON for the entire 24 hours and commitment of other units changes to achieve the optimal operation. The operation cost is calculated as $11262.8.

The comparison of distribution feeder net load with and without considering ramping constraint shows that in the case which there is no collaboration between the microgrid and the utility grid, i.e. no ramping constraints, even sharper ramps should be addressed by the utility grid. Indeed, in this case the microgrid exacerbated the distribution feeder net load variability, which should be supplied by the utility grid. The results show that in the absence of microgrid, the utility grid should address a maximum of 6 MW/h load change, or an average of 4.6 MW/h in 3 hours, while adding the microgrid in the feeder without consideration of any ramping constraints increases this amount to a maximum of 11.85 MW/h, or an average of 7 MW/h in 3 hours.

It is worthwhile to mention that the microgrid operation cost without consideration of any flexibility constraint is $11262.8, while it would be increased to $12126.3 after the addition of the ramping constraint. The reason of this cost increase, which should be paid to the microgrid by the utility grid, is the additional constraint that is imposed to the microgrid scheduling problem.

**Case 2:** The flexibility-oriented microgrid optimal scheduling considering prevailing uncertainties is solved for a 24-hour horizon. Forecast errors in distribution feeder load, solar generation, and market prices are considered as ±10%, ±20%, and ±10%, respectively. Furthermore, a 12-hour/day budget is considered as a limitation on uncertainty. A ramping limit of 2 MW/h is considered similar to Case 1.

Fig. 2 depicts the distribution feeder net load with and without considering flexibility constraint for ±10% load forecast error. As this figure shows, the utility grid encounters an average of 8.1 MW/h in 3 hours load change between hours 9 and 12 (maximum of 13.55 MW/h), as well as an average of 7.3 MW/h in 2 hours between hours 18 and 19 (maximum of 8 MW/h). The microgrid, however, restricts the ramping of the distribution feeder net load to 2 MW/h which has been requested by the utility grid. To obtain the desired ramping, the microgrid needs to deviate from its optimal schedule which leads to a $1652.9 increase in its operation cost. This 14.7% increase in the microgrid operation cost should be paid by the electric utility as an incentive for contribution in mitigating the net load ramping.

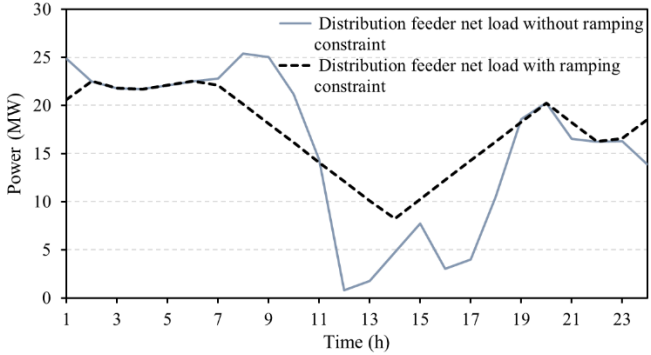

Fig. 2. Distribution feeder net load, with and without ramping constraint, for uncertain distributed load over the 24-hour scheduling horizon.

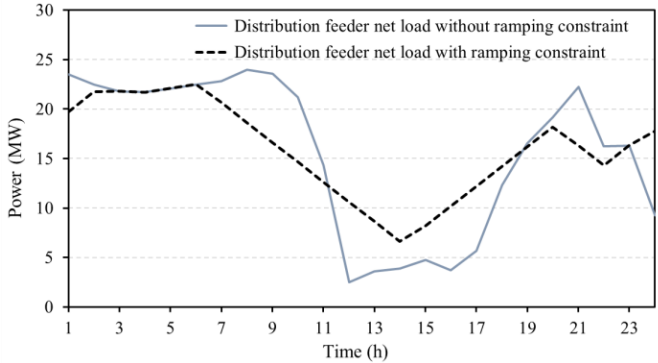

Fig. 3. Distribution feeder net load, with and without ramping constraint, for uncertain market price over the 24-hour scheduling horizon.

The obtained results for ±20% solar generation forecast error are almost the same as ±10% load forecast error with minor differences. For ±20% solar generation uncertainty, without any contribution from the microgrid, the utility grid encounters an average of 7.9 MW/h net load change in 3 hours between hours 10 and 12 (with a maximum of 15.75 MW/h), as well as an average of 6.8 MW/h in 2 hours between hours 18 and 19 (a maximum of 7.6 MW/h). The microgrid operation cost when capturing these ramps increases from $11262.8 to $12,642.2. The results show that although the load forecast error is ±10% compared with ±20% solar generation forecast error, the microgrid operation cost due to contribution in capturing ramping, for load uncertainty has been increased 2.5% more than solar generation uncertainty.

Fig. 3 demonstrates the distribution feeder net load with and without considering flexibility constraint for ±10% market price uncertainty. This figure shows the effectiveness of microgrid to address the distribution feeder net load. An average of 7 MW/h ramping in 3 hours in the morning, 11.85 MW/h load change between hours 11 and 12, and 6 MW ramping in one hour between hours 21 and 22 have been mitigated by the microgrid. The obtained results show that the microgrid operation cost is increased by $1516.7, equal to 13.3%, due to the addition of the flexibility constraint. It should be noted that in all cases unit 1 is ON for the entire scheduling horizon, while there are changes in commitment and dispatch of other units.

Table III summarizes the microgrid operation cost for studied cases. It clearly shows that considering uncertainty increases the microgrid operation cost, however it would be able to capture any possible deviations from the forecasted values. The table moreover shows the impact of different uncertainties on the microgrid operation cost.

TABLE III
MICROGRID OPERATION COST ($) FOR VARIOUS OPERATION SCHEDULING AND 2 MW/HOUR RAMPING LIMITS

| Microgrid optimal scheduling | Distributed load uncertainty | Distributed solar uncertainty | Market prices uncertainty |
|---|---|---|---|
| Ignoring uncertainty | $12126.3 | $12126.3 | $12126.3 |
| Considering uncertainty | $12915.7 | $12642.15 | $12862.9 |

**Case 3:** In this case the microgrid operation cost for various amounts of uncertainty budget are calculated. The obtained results in Table IV illustrate that the microgrid operation cost is directly proportional to the budget of uncertainty. The results further demonstrate that the changes on the load and solar generation have the highest and lowest impact on the microgrid operation cost, respectively. With increasing the budget of uncertainty 0 to 12 hours, the microgrid operation cost is increased by 6.5%, 4.25%, and 6% for distributed load, distributed solar generation, and market price uncertainty, respectively. It should be noted that ±10% was considered for load forecast error and market price uncertainty, whereas ±20% was considered for solar generation uncertainty.

TABLE IV
MICROGRID OPERATION COST FOR VARIOUS BUDGETS OF UNCERTAINTY (CONSIDERING A 2 MW/HOUR RAMPING LIMIT)

| Budget of Uncertainty (h) | Distributed load uncertainty | Distributed solar uncertainty | Market price uncertainty |
|---|---|---|---|
| 0 | $12,126.3 | $12,126.3 | $12,126.3 |
| 3 | $12,526.6 | $12,393.4 | $12,446.1 |
| 6 | $12,715.3 | $12,561.9 | $12,611.7 |
| 9 | $12,850.1 | $12,607.9 | $12,748.6 |
| 12 | $12,915.7 | $12,642.2 | $12,862.9 |

IV. CONCLUSIONS

A flexibility-oriented microgrid optimal scheduling under uncertainty was proposed in this paper to address distribution network net load ramping. The robust optimization method was used for capturing uncertainties and increasing the practicality of the proposed model. The obtained results showed that utilizing the microgrid decreases the utility ramping to the desired amounts. Although flexibility constraints led to higher microgrid operation cost, which should be paid to the microgrid by the electric utility, it removed the need for costly investments on reinforcing the existing electricity infrastructure. The numerical simulations further showed that by increasing the budget of uncertainty, the microgrid operation cost increases as it was required to capture uncertainty in a larger number of hours. In addition, the obtained results indicated that the microgrid operation cost is more sensitive to load uncertainty compared to renewable generation and price uncertainty.